\documentclass[sigconf,natbib=true,nonacm=true]{acmart}

\usepackage[linesnumbered,boxed,ruled]{algorithm2e}
\usepackage{multirow}
\usepackage[most]{tcolorbox}
\AtBeginDocument{%
  }

\begin{document}

%%
%% The "title" command has an optional parameter,
%% allowing the author to define a "short title" to be used in page headers.
\title{CAST: Modeling Semantic-Level Transitions for Complementary-Aware Sequential Recommendation}
% Complementary Next-Item Prediction Beyond Co-occurrence via Semantic Transition Priors

%%
%% The "author" command and its associated commands are used to define
%% the authors and their affiliations.
%% Of note is the shared affiliation of the first two authors, and the
%% "authornote" and "authornotemark" commands
%% used to denote shared contribution to the research.
\author{Qian Zhang}
\email{zhaqi075@student.otago.ac.nz}
\orcid{0009-0009-1117-0236}
\affiliation{%
  \institution{School of Computing, University of Otago}
  \city{Dunedin}
  \country{New Zealand}
}
\author{Lech Szymanski}
\email{lech.szymanski@otago.ac.nz}
\orcid{0000-0002-5192-0304}
\affiliation{%
  \institution{School of Computing, University of Otago}
  \city{Dunedin}
  \country{New Zealand}
}
\author{Haibo Zhang}
\email{haibo.zhang@unsw.edu.au}
\orcid{0000-0002-3752-0806}
\affiliation{%
  \institution{University of New South Wales}
  \city{Sydney}
  \country{Australia}
}
\author{Jeremiah D. Deng}
\email{jeremiah.deng@otago.ac.nz}
\orcid{0000-0003-3727-4403}
\affiliation{%
  \institution{School of Computing, University of Otago}
  \city{Dunedin}
  \country{New Zealand}
}

%%
%% By default, the full list of authors will be used in the page
%% headers. Often, this list is too long, and will overlap
%% other information printed in the page headers. This command allows
%% the author to define a more concise list
%% of authors' names for this purpose.
\renewcommand{\shortauthors}{Zhang et al.}

%%
%% The abstract is a short summary of the work to be presented in the
%% article.
\begin{abstract}
  Sequential Recommendation (SR) aims to predict the next interaction of a user based on their behavior sequence, where complementary relations often provide essential signals for predicting the next item. However, mainstream models relying on sparse co-purchase statistics often mistake spurious correlations (e.g., due to popularity bias) for true complementary relations. Identifying true complementary relations requires capturing the fine-grained item semantics (e.g., specifications) that simple cooccurrence statistics would be unable to model. While recent semantics-based methods utilize discrete semantic codes to represent items, they typically \textbf{aggregate} semantic codes into coarse item representations. This aggregation process blurs specific semantic details required to identify complementarity. To address these critical limitations and effectively leverage semantics for capturing reliable complementary relations, we propose a \textbf{C}omplementary-\textbf{A}ware \textbf{S}emantic \textbf{T}ransition (\textbf{CAST}) framework that introduces a new modeling paradigm built upon \textbf{semantic-level} transitions. Specifically, a semantic-level transition module is designed to model dynamic transitions directly in the discrete semantic code space, effectively capturing fine-grained semantic dependencies often lost in aggregated item representations. Then, a complementary prior injection module is designed to incorporate LLM-verified complementary priors into the attention mechanism, thereby prioritizing complementary patterns over co-occurrence statistics. Experiments on multiple e-commerce datasets demonstrate that CAST consistently outperforms the state-of-the-art approaches, achieving up to 17.6\% Recall and 16.0\% NDCG gains with 65$\times$ training acceleration. This validates its effectiveness and efficiency in uncovering latent item complementarity beyond statistics. The code will be released upon acceptance.
\end{abstract}

%%
%% The code below is generated by the tool at http://dl.acm.org/ccs.cfm.
%% Please copy and paste the code instead of the example below.
%%


%%
%% Keywords. The author(s) should pick words that accurately describe
%% the work being presented. Separate the keywords with commas.

%\received{20 February 2007}
%\received[revised]{12 March 2009}
%\received[accepted]{5 June 2009}

%%
%% This command processes the author and affiliation and title
%% information and builds the first part of the formatted document.
\maketitle

\section{Introduction}
Sequential recommendation (SR) aims to predict the next interaction of a user by modeling the temporal dynamics of historical behaviors \cite{kang2018self,sun2019bert4rec,du2023frequency}. However, several factors complicate accurate prediction: user sequences may span multiple months, and purchase decisions are often driven by the intrinsic semantics of items that determine their compatibility. Furthermore, users interact with relatively few items from the large-scale item set, and the recorded sequences often diverge from true user preferences due to accidental clicks, naturally resulting in severe data sparsity and noise~\cite{xie2022contrastive,zhou2024contrastive}. In such sparse and noisy environments, purely statistical signals (e.g., co-purchase frequency) become unreliable, as they frequently conflate accidental co-occurrences with genuine user interests. To robustly model user intent under these conditions, it is crucial to leverage signals that are invariant to statistical noise: \textit{the intrinsic functional relationships between items, specifically complementary relations} (e.g., Camera $\rightarrow$ SD Card). Research \cite{mcauley2015inferring,wang2020make} has demonstrated that complementary relations play a crucial role in next-item prediction, as they capture user intent rather than merely the preference for similar items. Incorporating these relations allows models to transcend simple temporal adjacency and uncover the \textit{underlying reasons} that drive item transitions~\cite{zhang2025semantic}.

\begin{figure}[t]
  \centering
  \includegraphics[width=1.0\linewidth]{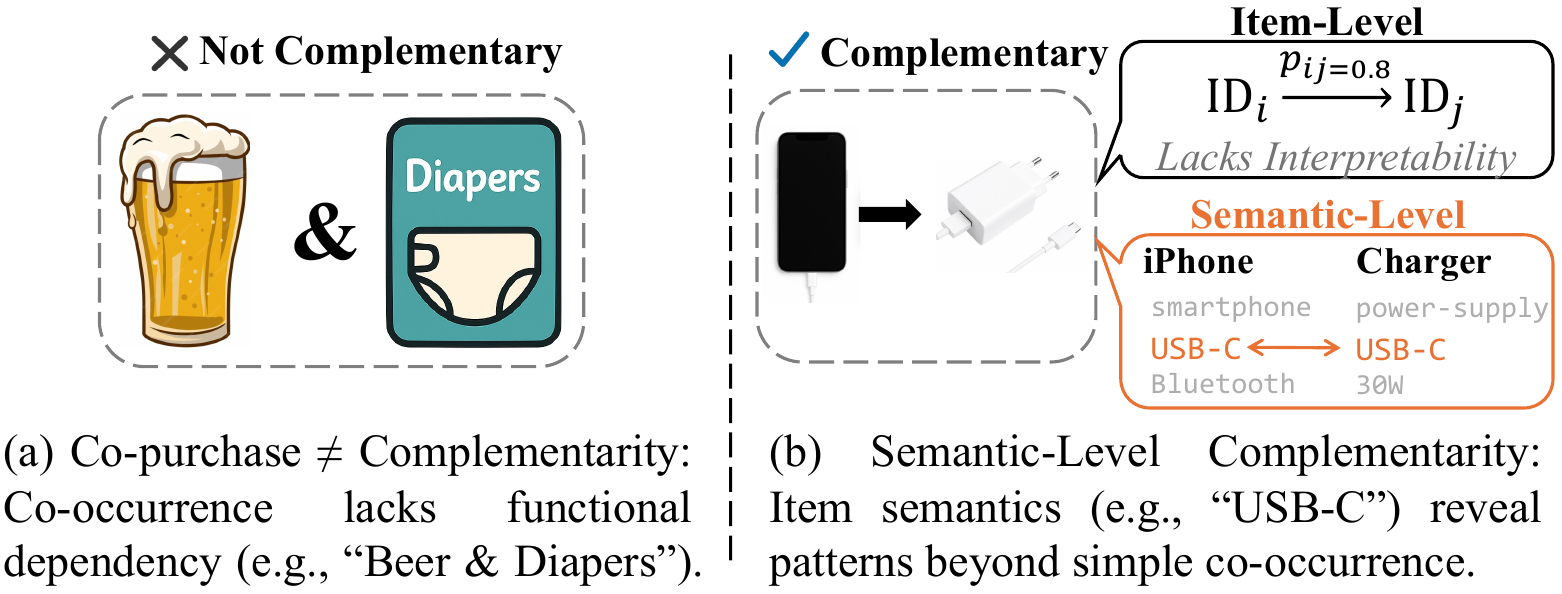}
  \caption{Example of why co-purchase does not equal complementarity and why semantic-level modeling is needed.}
  \label{fig:intro}
\end{figure}

However, existing approaches generally infer complementary relations from co-purchase frequency \cite{zhang2021learning,yan2022personalized}. Such statistical signals capture mere correlations rather than reliable functional dependencies. Consequently, they are susceptible to external factors like exposure bias and seasonality~\cite{schnabel2016recommendations,zhang2021causal}: \textbf{items frequently co-purchased are not necessarily complementary}. The classic ``beer and diapers'' retail case illustrates this pitfall: such co-purchase patterns may stem from distinct user intentions rather than reflecting true functional complementarity, as described in Figure~\ref{fig:intro}(a). Relying solely on these unstable statistical artifacts risks learning spurious relations that generalize poorly to sparse or cold-start scenarios, where functional logic is required to infer user intent, not just historical frequency.

To distinguish true complementary relations from spurious co-purchases, it is essential to look beyond behavioral statistics and ground recommendations in \textbf{item semantics}, which are crucial for functional compatibility. However, mainstream sequential recommendation models (SASRec~\cite{kang2018self}, FEARec~\cite{du2023frequency}, etc.) treat items as discrete ID embeddings, lacking the semantic rationale behind item transitions. Although recent semantic-aware methods (e.g., VQ-Rec~\cite{hou2023learning}) have introduced semantic codes to capture textual information, they typically aggregate semantic codes into a coarse item representation before sequence modeling. This ``aggregation-then-modeling'' paradigm restricts semantic-aware methods to item-level transitions, masking the specific semantic attributes (e.g., a "USB-C" interface) that actually drive compatibility. As illustrated in Figure~\ref{fig:intro}(b), complementarity is often driven by the matching of specific attributes rather than whole items. Therefore, effectively modeling complementary relations requires shifting the focus from item-level transitions to \textbf{fine-grained semantic transitions}.

To address these limitations, we propose \textbf{CAST}, a \textbf{C}omplementary-\textbf{A}ware \textbf{S}emantic \textbf{T}ransition framework for sequential recommendation. Specifically, we encode the textual features of items with a pre-trained language model and discretize the representations into a shared semantic codebook, so each item is represented as a sequence of semantic codes. Instead of modeling item-level transitions, CAST explicitly models \emph{code-level} semantic transitions and learns a dynamic transition matrix in the semantic code space to capture functional complementarity beyond spurious co-occurrence (statistical correlation). To mitigate noisy co-purchase signals, we construct a complementary transition prior from complementary relations verified with item texts (e.g., with LLM assistance) and inject it as an additive bias into self-attention, guiding the model toward plausible semantic transitions even under sparsity and noise.

The main contributions of this work are summarized as follows:
\begin{itemize}
  \item We propose \textbf{CAST}, which shifts the modeling paradigm from item-level to \textit{fine-grained code-level transitions}. By explicitly learning dynamic dependencies in the discrete semantic space, CAST captures compatibility patterns beyond static co-occurrence statistics.
  \item We develop a \textbf{subspace-preserving alignment} strategy to maintain the orthogonality of semantic codes, and a \textbf{transition-guided self-attention} mechanism that incorporates LLM-verified complementary priors as adaptive transition biases to guide semantic dependencies.
  \item Extensive experiments demonstrate that CAST consistently outperforms state-of-the-art baselines. Further analysis verifies its ability to reveal distinct complementary patterns, while achieving over 65$\times$ training acceleration compared to the strongest baseline.
\end{itemize}

\section{Task Formulation}
\label{sec:formulation}
Let $\mathcal{U}$ and $\mathcal{V}$ denote the sets of users and items, where $|\mathcal{U}|$ and $|\mathcal{V}|$ are the respective numbers of users and items.  $\mathcal{X} = \{x_v\}_{v \in \mathcal{V}}$ denotes the set of textual features (e.g., titles, categories) associated with the items. Each user has an interaction sequence $s=[v_1, v_2, \ldots, v_{n}]$, where $v_t \in \mathcal{V}$ is the item interacted with at the time step $t$, and $n$ is the length of the sequence. The sequential recommendation task aims to predict the next item with which a user will interact, given the observed history up to $t$ ($1 < t < |s|$). During training, the model takes the first $t$ items $(v_1, v_2, \ldots, v_t)$ as input and learns to predict the next item $v_{t+1}$.

\section{Methodology}
\begin{figure*}[h]
  \centering
  \includegraphics[width=0.85\textwidth]{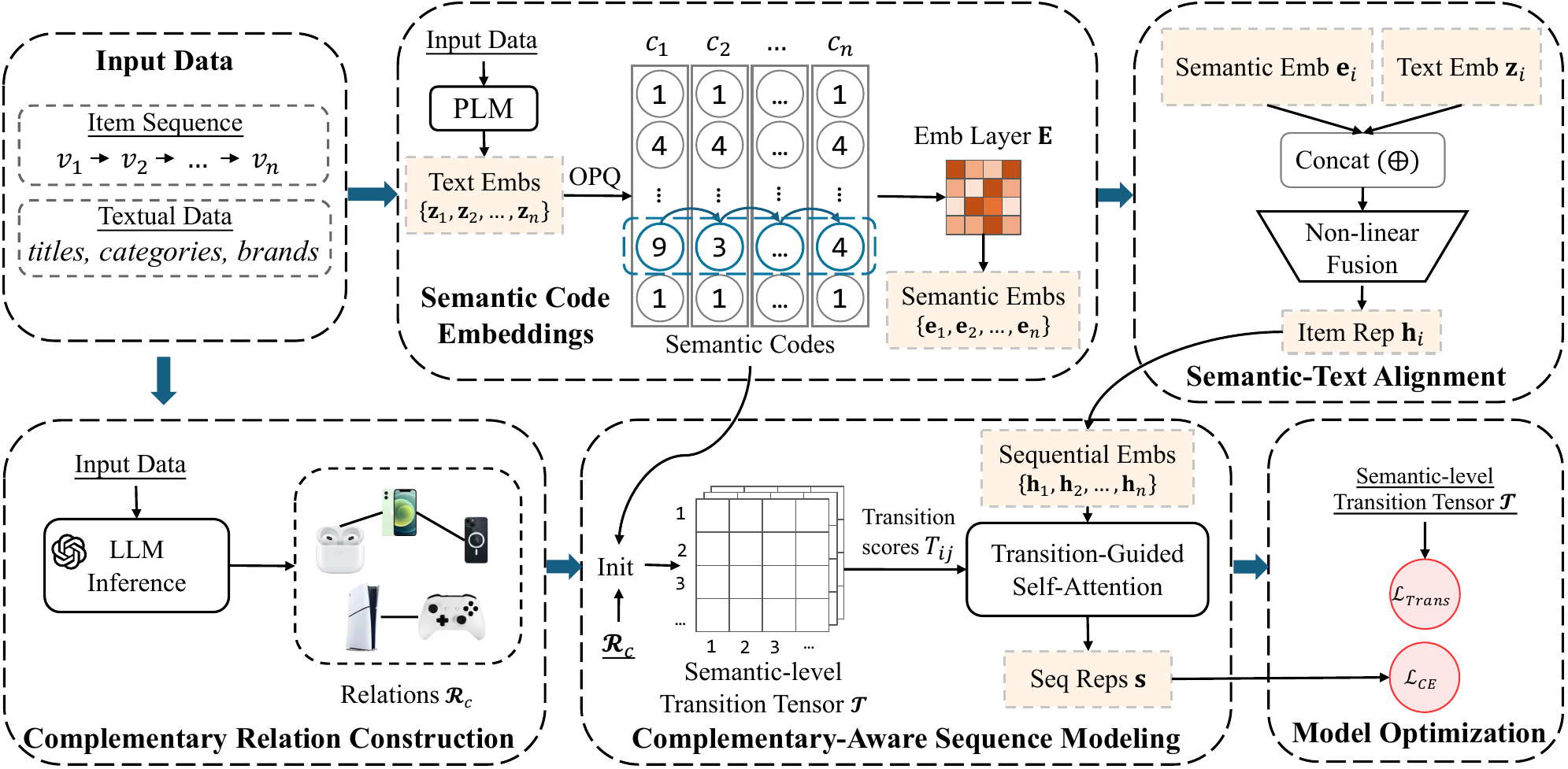}
  \caption{Overview of the CAST framework. CAST integrates LLM-verified complementary relations ($\mathcal{R}_c$) and OPQ-based semantic codes. First, the \textit{Semantic-Text Alignment} module aligns semantic code embeddings with textual embeddings to generate item representations ($\mathbf{h}_i$). Subsequently, the \textit{Complementary-Aware Sequence Modeling} module utilizes a semantic-level transition tensor ($\mathcal{T}$) guided by the complementary priors to modulate the self-attention mechanism. Finally, the model is optimized using a joint objective combining recommendation ($\mathcal{L}_{CE}$) and transition consistency ($\mathcal{L}_{Trans}$) losses.}
  \label{fig:model}
\end{figure*}

The overall architecture of the proposed CAST framework is illustrated in Figure~\ref{fig:model}. Generally, CAST consists of four core components: (1) \textbf{Complementary Relation Construction} (Section \ref{sec:comp-construction}), which leverages LLM-based verification to construct reliable complementary relations, filtering spurious co-occurrences from intrinsic functional dependencies; (2) \textbf{Semantic Code Embeddings and Alignment} (Sections \ref{sec:sem-embedding} and \ref{sec:alignment}), which discretizes item textual features into fine-grained semantic codes via Optimized Product Quantization (OPQ) and aligns them with textual semantics through a subspace-preserving MLP projector; (3) \textbf{Complementary-Aware Sequence Modeling} (Section \ref{sec:comp-modeling}), which explicitly captures the dynamic evolution of user interests by injecting LLM-verified complementary priors into the self-attention mechanism via a learnable semantic transition tensor; and (4) \textbf{Model Optimization} (Section \ref{sec:optimization}), which trains the model end-to-end using a next-item prediction objective augmented by a transition consistency regularization to ensure the reliability of learned semantic dependencies.

\begin{algorithm}[t]
\caption{Complementary Relation Construction using LLM}
\label{algorithm:construction}
  \SetAlgoLined 
  \DontPrintSemicolon
  \KwIn{Interaction sequences $\mathcal{S} = \{s_1, s_2, \dots, s_{|\mathcal{S}|}\}$, item texts $\mathcal{X} = \{x_v\}_{v \in \mathcal{V}}$, window size $w$, frequency threshold $\theta_f$, LLM complementary confidence threshold $\theta_c$, textual similarity threshold $\theta_s$}
  \KwResult{Complementary relation set $\mathcal{R}_c$}

%    \State \textbf{Complementary Candidate Generation from Co-purchase Data}
	\tcc{Obtain complementary candidates from co-purchases}
	
         \ForAll{$s = [v_1, v_2, \dots, v_n] \in \mathcal{S}$} {
               $\mathrm{freq}(v_i, v_j) \gets \mathrm{freq}(v_i, v_j) + 1, \quad \forall i < j, j - i < w$ 
               \tcp{Count co-purchases}
            }
           Initialize \textit{raw} complementary relation set $\mathcal{R}_c^{0} \gets \{ (v_i,v_j) \mid \mathrm{freq}(v_i,v_j) \ge \theta_f \}$ \;
	\tcc{Complementary inference via LLM}
	Initialize complementary relation set $\mathcal{R}_c \gets \emptyset$\;
	\ForAll{$(v_i, v_j) \in \mathcal{R}_c^0$} {
                    $w_{ij} \gets \mathrm{CompScore}(x_i, x_j)$ \tcp{Query LLM with prompt for complementary confidence score} 
                    \If{$w_{ij}\ge \theta_c$} {
                        $\mathcal{R}_c \gets \mathcal{R}_c \cup (v_i, v_j, w_{ij})$ 
                        \tcp{Add in }
                    }
                }
	\tcc{Get substitutable relations by similarity}
	Initialize substitutable relation set $\mathcal{R}_s \gets \emptyset$\;
	 \ForAll{\rm{item pairs} $(v_i, v_j)$} {
                \If{$\mathrm{sim}(\mathrm{PLM}(x_i), \mathrm{PLM}(x_j)) \geq \theta_s$} {
                    $\mathcal{R}_s \gets \mathcal{R}_s \cup (v_i, v_j)$ \tcp{substitutable}
		}
	}
	\tcc{Expand complementary relations}
	\ForAll{$(v_i,v_j,w_{ij}) \in \mathcal{R}_c$} {
                \ForAll{$v_k: (v_i,v_k)\in \mathcal{R}_s$} {
                    $\mathcal{R}_c \gets \mathcal{R}_c \cup (v_k,v_j,w_{ij}) $
                }
          }
          \tcc{Symmetrize $\mathcal{R}_c$}
          \ForAll{$(v_i,v_j,w_{ij}) \in \mathcal{R}_c$} {
			$\mathcal{R}_c \gets \mathcal{R}_c \cup (v_j,v_i,w_{ij})$, {\bf if} $(v_j,v_i,w_{ij}) \notin \mathcal{R}_c$
	}

\Return $\mathcal{R}_c$
\end{algorithm}

\begin{figure}[t]
\centering
\begin{tcolorbox}[
    colback=gray!5, 
    colframe=black, 
    rounded corners, 
    boxrule=0.8pt, 
    fontupper=\normalsize,
    left=2mm, right=2mm, top=2mm, bottom=2mm
]
    \textbf{Task Description:} You are an assistant who determines to what extent two products are complementary on a [0, 1] scale.
    
    \textbf{Evaluation Criteria:}
    \begin{enumerate}
        \item \textit{Direct Interaction:} Are they often used together for the same intent?
        \item \textit{Functional Enhancement:} Does one enhance the functionality of the other?
        \item \textit{Market Relationship:} Considerations of market co-occurrence.
    \end{enumerate}
    
    \textbf{Input:}
    \begin{itemize}
    \item Product 1: [Item Title, Categories, Brand]
    \item Product 2: [Item Title, Categories, Brand]
    \end{itemize}
    
    \textbf{Output Requirements:}
    \begin{enumerate}
    \item Step-by-step reasoning referencing the criteria.
    \item A single numeric score from 0 to 1.
    \end{enumerate}
\end{tcolorbox}
\caption{The prompt template utilized for inferring complementary relations via LLM.}
\label{fig:prompt}
\end{figure}

\subsection{Complementary Relation Construction}
\label{sec:comp-construction}
We construct the complementary relation set $\mathcal{R}_c$ following Algorithm \ref{algorithm:construction}. First, we generate raw candidates ($\mathcal{R}_c^{0}$) from co-purchase data using a sliding window $w$ and a frequency threshold $\theta_f$ (Lines 1--4). Second, for each candidate pair $(v_i, v_j)$, an LLM evaluates the two items based on their textual metadata (e.g., title) and outputs a complementary confidence score $w_{ij}\in[0,1]$ (higher value means complementary to a higher degree). The prompt template used for this evaluation is detailed in Figure \ref{fig:prompt}, designed to focus on direct interaction and functional enhancement (Line 7). We keep pairs with $w_{ij}\ge \theta_c$ and form the initial set $\mathcal{R}_c=\{(v_i, v_j, w_{ij})\}$ (Lines 8--9). Third, we build an auxiliary substitutable set $\mathcal{R}_s$ containing item pairs with high textual similarity ($\ge \theta_s$) (Lines 14--15). Finally, we expand $\mathcal{R}_c$ using a transitive rule (Lines 18--22): if $(v_i, v_j, w_{ij}) \in \mathcal{R}_c$ and substitutable pair $(v_i, v_k) \in \mathcal{R}_s$, then the pair $(v_j, v_k, w_{ij})$ is added to $\mathcal{R}_c$ (i.e., we propagate the weight $w_{ij}$ from $(v_i,v_j)$). Subsequently, $\mathcal{R}_c$ is used to inform the model about the transition probabilities between semantic tokens, thereby embedding complementary knowledge directly in item representations. Specific parameter settings and their justifications are detailed in Section \ref{sec:implementation_details}.

\subsection{Semantic Code Embeddings}
\label{sec:sem-embedding}

Recent works on vector-quantized item representations~\cite{jegou2010product,hou2023learning,liu2025bridging} demonstrate that discrete semantic coding effectively captures item semantics. Following this paradigm, we adopt \textit{optimized product quantization (OPQ)} \cite{ge2013optimized}, which discretizes the textual embedding of each item in a sequence of discrete semantic codes that serve as generalizable semantic representations. Given an item $v_i$ with textual feature $x_i$ we obtain its textual embedding $\hat{\mathbf{z}}_i \in \mathbb{R}^{d_{\text{text}}}$ via a frozen pre-trained language model (PLM) encoder.
\begin{equation}
    \hat{\mathbf{z}}_i = \text{PLM}(x_i).
\end{equation}

Subsequently, the OPQ method discretizes $\hat{\mathbf{z}}_i$ into an item-specific code sequence $\mathbf{c}_i = (c_i^{(1)}, \ldots, c_i^{(D)})$, corresponding to the column vectors in the ``Semantic Codes'' block of Figure~\ref{fig:model} (where the figure depicts a sequence of $n$ items). Concretely, OPQ partitions $\hat{\mathbf{z}}_i$ into $D$ disjoint sub-vectors and quantizes each sub-vector to its nearest centroid in a codebook of size $C$ (where $D$ and $C$ denote the number of quantized subspaces and the codebook size, respectively). For instance, if $D=3$, an item (e.g., ``Wireless Mouse'') might be encoded as $\mathbf{c}_i = (12, 245, 5)$. Here, each code represents a \textit{latent semantic unit} derived from clustering, capturing specific semantic patterns (e.g., functional attributes) in the subspace. Formally, this yields a discrete code $c_i^{(k)} \in \{1, 2, \ldots, C\}$ for each subspace $k \in \{1,\ldots,D\}$.
%where $c^{(k)} \in \{1, 2, \ldots, C\}$ for each subspace $k = (1, \ldots, D)$, and $C$ is the number of representative centroids per subspace, $D$ denotes the number of quantized subspaces.

Based on these learned discrete codes, we construct a set of $D$ semantic code embedding matrices $\{\mathbf{E}^{(1)}, \ldots, \mathbf{E}^{(D)}\}$, where each $\mathbf{E}^{(k)} \in \mathbb{R}^{C \times d}$ (with $d$ being the embedding dimension) is shared across all items and the $m$-th row $\mathbf{E}^{(k)}[m]$ stores the learnable embedding of centroid $m$ in the $k$-th subspace. 

Given the code sequence $\mathbf{c}_i = (c_i^{(1)}, \ldots, c_i^{(D)})$ of an item, we retrieve its subspace-wise semantic embeddings by lookup:
\begin{equation}
    \mathbf{e}_i^{(k)} = \mathbf{E}^{(k)}[c_i^{(k)}], \quad k=1,\ldots,D.
\end{equation}
We denote the retrieved sequence of subspace embeddings for item $v_i$ as $\mathbf{e}_i = (\mathbf{e}_i^{(1)}, \ldots, \mathbf{e}_i^{(D)})$, which serves as its semantic representation in the subsequent modules.

\subsection{Subspace-Preserving Semantic-Text Alignment}
\label{sec:alignment}

Built on OPQ, the generated semantic IDs are decomposed so that tokens $[c^{(1)}, c^{(2)}, ..., c^{(D)}]$ at different positions correspond to disjoint quantized subspaces (one codebook per position). However, standard aggregation techniques (e.g., \textit{mean pooling} or \textit{attention}) will mix these subspace-specific positions, blurring the orthogonality of the subspace codebooks and hindering the fine-grained alignment from text semantics to distinct subspaces. To address this, we propose a \textbf{subspace-preserving semantic-text alignment} strategy. We first preserve the positional independence of subspaces by flattening the semantic embeddings for item $v_i$: $\mathbf{h}^{\text{flat}}_{i} = \mathbf{e}_{i}^{(1)} \oplus \mathbf{e}_{i}^{(2)} \oplus \cdots \oplus \mathbf{e}_{i}^{(D)} \in \mathbb{R}^{D \cdot d}$, where $\oplus$ denotes the vector concatenation operation. To align the dimensionality, we first project the textual embedding $\hat{\mathbf{z}}_i$ to the hidden dimension $d$ via a learnable linear projection $\mathbf{W}_{\text{proj}} \in \mathbb{R}^{d \times d_{\text{text}}}$:
\begin{equation}
    \mathbf{z}_i = \mathbf{W}_{\text{proj}}\hat{\mathbf{z}}_i \in \mathbb{R}^{d}.
\end{equation}

To bridge the gap between the disjoint semantic subspaces (each corresponding to one quantized codebook indexed by $k \in \{1,\ldots,D\}$) and the textual representations, we employ a Multi-Layer Perceptron (MLP) as a non-linear fusion projector. Drawing inspiration from the efficacy of non-linear transformations in representation learning \cite{chen2020simple}, we map the composite features into a unified latent space via a dimensionality-reduction architecture:
\begin{equation}
\label{eq:nonlinear}
    \mathbf{h}_i = \text{MLP}(\mathbf{h}_{i}^{\text{flat}} \oplus \mathbf{z}_i) = \mathbf{W}_2\,\phi_\eta(\mathbf{W}_1(\mathbf{h}_{i}^{\text{flat}} \oplus \mathbf{z}_i) + \mathbf{b}_1) + \mathbf{b}_2,
\end{equation}
where $\phi_\eta(\cdot)$ denotes the function consisting of the GeLU activation followed by a dropout operation with probability $\eta$. $\mathbf{W}_1 \in \mathbb{R}^{d' \times (D+1)d}$ and $\mathbf{W}_2 \in \mathbb{R}^{d \times d'}$ are learnable weight matrices (with hidden size $d'$), and $\mathbf{b}_1, \mathbf{b}_2$ are learnable bias vectors. This module serves two functions: (1) \textbf{Non-linear Mapping}: it learns the specific correlations between the text and each semantic subspace, mapping specific text semantics to the corresponding codebook representations; and (2) \textbf{Dimensionality Reduction}: it compresses the high-dimensional flattened input ($D \cdot d$) back to the hidden dimension ($d$), effectively removing redundancy while retaining the essential semantic signals aligned with the textual context. Notably, this MLP-based method significantly improves the efficiency of the alignment task, replacing conventional methods like cross-attention ($\mathcal{O}(N^2)$)~\cite{li2023blip} with a linear MLP projection ($\mathcal{O}(N)$). This lightweight design substantially reduces computational overhead without compromising alignment efficacy (as validated in Section \ref{sec:efficiency}), making the model scalable for large-scale candidate sets.

\subsection{Complementary-Aware Sequence Modeling}
\label{sec:comp-modeling}
Conventional SRs typically rely on self-attention mechanisms to capture local sequential dependencies but ignore global complementary relations absent from the current context. To bridge this gap, we propose a complementary-aware sequence modeling framework. Instead of static injection, CAST establishes a mutually reinforcing mechanism, where the global transition matrix $\mathcal{T}$ serves as a learnable bias to guide attention distribution, while updating $\mathcal{T}$ through training to learn refined semantic transition patterns.

\subsubsection{Semantic-Level Transition Construction}
To model the complementary relations in the latent semantic space rather than the sparse item space, we define a semantic-level complementary transition tensor $\mathcal{T} \in \mathbb{R}^{D \times C \times C}$, where $D$ corresponds to the number of semantic subspaces (as defined in Section \ref{sec:sem-embedding}), and $C \times C$ represents the transition matrix between semantic tokens in each semantic subspace. Specifically, $\mathcal{T}_{k, c, c^{\prime}}$ represents the transition score from semantic token $c$ to token $c^{\prime}$ in the $k$-th subspace.

\textbf{Prior Initialization via Complementary Relations}.
To inject high-quality relation priors and accelerate convergence, we initialize $\mathcal{T}$ using the complementary relation set $\mathcal{R}_c$ constructed in Algorithm \ref{algorithm:construction}. Since $\mathcal{R}_c$ is defined at the item-level while $\mathcal{T}$ models transitions between semantic tokens, we need to map complementary item pairs into subspace-specific token transitions. Specifically, for each complementary item pair $(v_i, v_j)$ with complementary weight $w_{ij}$ in $\mathcal{R}_c$ (LLM-derived complementarity confidence), we retrieve their corresponding semantic token sequences $\mathbf{c}_i = (c_i^{(1)}, \dots, c_i^{(D)}),\, \mathbf{c}_j = (c_j^{(1)}, \dots, c_j^{(D)}) \in \{1,\dots,C\}^{D}$. For each semantic subspace $k \in \{1,\dots,D\}$, we construct a token-level co-occurrence matrix $\mathbf{M}_k \in \mathbb{R}^{C \times C}$ initialized to zeros. For every complementary pair $(v_i, v_j, w_{ij}) \in \mathcal{R}_c$, we update:
\begin{equation}
\label{eq:comp-matrix}
\mathbf{M}_k\big[c_i^{(k)},\, c_j^{(k)}\big] \;\leftarrow\;
\mathbf{M}_k\big[c_i^{(k)},\, c_j^{(k)}\big] + w_{ij},
\qquad \forall\, k = 1,\dots,D.
\end{equation}

Since complementary relations are undirected, we symmetrize the subspace-specific matrix,
\begin{equation}
\widetilde{\mathbf{M}}_k
= \frac{1}{2}\big(\mathbf{M}_k + \mathbf{M}_k^\top\big).
\end{equation}
To avoid numerical issues caused by zero counts, we add a small smoothing constant $\epsilon$ and transform the counts into transition logits,
\begin{equation}
\mathcal{T}_{k, c_i^{(k)}, c_j^{(k)}}^{(0)}
= \log\big(\mathbf{M}_k\big[c_i^{(k)},\, c_j^{(k)}\big] + \epsilon\big),
\end{equation}
used to initialize the learnable transition tensor $\mathcal{T}$. During training, $\mathcal{T}$ is further updated end-to-end together with other parameters.

\subsubsection{Transition-Guided Self-Attention}

To incorporate the learned complementary transition prior into sequence modeling, we inject $\mathcal{T}$ into the self-attention mechanism. At initialization and during training, we standardize the transition scores to stabilize optimization, obtaining $\mathbf{P}_k = \operatorname{Standardize}(\mathcal{T}_k)$, where values in $\mathbf{P}_k$ represent the Z-score normalized transition logits between semantic tokens in subspace $k$. This ensures that the transition priors are on a scale comparable to the attention logits.

We then modify the self-attention mechanism by injecting a transition score into the attention logits. Given the hidden state $\mathbf{h}_i$ of item $v_i$ in user sequence, a self-attention head computes its query, key, and value as $\mathbf{Q}_i = \mathbf{W}_\text{Q} \mathbf{h}_i$, $\mathbf{K}_i = \mathbf{W}_\text{K} \mathbf{h}_i$, and $\mathbf{V}_i = \mathbf{W}_\text{V} \mathbf{h}_i$, where $\mathbf{W}_\text{Q}, \mathbf{W}_\text{K}, \mathbf{W}_\text{V} \in \mathbb{R}^{d \times d}$ are learnable projection matrices. To incorporate the complementary transition prior into the attention computation, we first define a semantic transition scoring function $T(v_a, v_b)$ that measures the transition strength from item $v_a$ to item $v_b$. Let $c_a^{(k)}$ and $c_b^{(k)}$ denote the semantic tokens of items $v_a$ and $v_b$ in subspace $k$. The score is calculated as:

\begin{equation}
\label{eq:trans-score}
T(v_a, v_b) = \sum_{k=1}^{D} \omega_k \cdot \mathbf{P}_k[c_a^{(k)}, c_b^{(k)}],
\end{equation}
where $\omega_k$ represents the learnable importance weight of subspace $k$ (normalized using Softmax). $\mathbf{P}_k$ contains standardized log-probabilities (scores).

In the self-attention mechanism, when attending from the current item $v_i$ (\textit{query}) to a historical item $v_j$ (\textit{key}), we inject the semantic transition bias directly into the attention logits. Note that the transition score $T(v_j, v_i)$ explicitly models the probability from the historical item $v_j$ to the current item $v_i$ (i.e., $v_j \rightarrow v_i$). By injecting this directional prior, we ensure that the attention mechanism prioritizes historical items that are functionally likely to lead to the current item. The semantic-aware attention weights are computed as:
\begin{equation}
\label{eq:trans-attn}
    \alpha_{i,j} = \frac{\exp\!\left( \frac{\mathbf{Q}_i^\top \mathbf{K}_j}{\sqrt{d}} + \lambda \cdot T(v_j, v_i) \right)}{\sum_{p=1}^{i} \exp\!\left( \frac{\mathbf{Q}_i^\top \mathbf{K}_p}{\sqrt{d}} + \lambda \cdot T(v_p, v_i) \right)},
\end{equation}
where $\alpha_{i,j}$ denotes the attention weight assigned to the historical item $v_j$ at position $j$ ($j \le i$) and $\sqrt{d}$ is the scaling factor based on the embedding dimension $d$. The scalar $\lambda$ controls the strength of the transition score $T(v_j, v_i)$.

Based on the attention weights $\alpha_{i,j}$ in Eq.~\eqref{eq:trans-attn}, each attention head computes its output as $\mathbf{o}_i = \sum_{j \le i} \alpha_{i,j} \mathbf{V}_j$. These weights $\alpha_{i,j}$ are exactly the attention coefficients used in the relation-aware multi-head attention ($\operatorname{MHA_{rel}}$). We stack $L$ layers of such relation-aware Transformer blocks to capture sequential dependencies. Formally, let $\mathbf{H}^{(0)} = [\mathbf{h}_1 + \mathbf{p}_1, \dots, \mathbf{h}_n + \mathbf{p}_n]$ denote the input sequence embeddings, where $\mathbf{h}_i$ represents the fused item embedding in Section \ref{sec:alignment} and $\mathbf{p}_i$ is the absolute position embedding. The hidden states are updated iteratively through the layers:

\begin{equation}
\mathbf{H}^{(l)} = \operatorname{FFN}\left( \operatorname{MHA_{rel}}\left( \mathbf{H}^{(l-1)} \right) \right), \quad l \in \{1, \dots, L\},
\end{equation}
where $\operatorname{MHA_{rel}}(\cdot)$ denotes the multi-head attention mechanism explicitly modulated by the semantic transition score $T(v_j, v_i)$, and $\operatorname{FFN}(\cdot)$ represents the feed-forward network. Following previous studies \cite{hou2023learning,liu2025bridging}, we utilize the hidden state of the last item at the final layer, denoted by $\mathbf{s} = \mathbf{h}_n^{(L)}$, as the overall sequence representation to compute the prediction scores with candidate items.

\subsection{Model Optimization}
\label{sec:optimization}
We employ a joint optimization strategy to train the model end-to-end, aiming to maximize next-item prediction accuracy while explicitly structuring the semantic transition patterns via auxiliary supervision.

\subsubsection{Next-item Prediction}
The primary objective is to maximize the likelihood of the ground-truth next item $v_{t+1}$ given the sequence representation $\mathbf{s}$. We employ the Cross-Entropy (CE) loss with softmax normalization:
\begin{equation}
\mathcal{L}_{\text{CE}} = - \log \frac{\exp(\mathbf{s}^\top \mathbf{h}_{v_{t+1}} / \tau)}{\sum_{v' \in \mathcal{V}} \exp(\mathbf{s}^\top \mathbf{h}_{v'} / \tau)},
\end{equation}
where $\mathbf{s}$ is the sequence representation (as defined in Section~\ref{sec:comp-modeling}), $\mathbf{h}_v$ denotes the item representation (as defined in Section~\ref{sec:alignment}), and $\tau$ is the temperature coefficient.

\subsubsection{Transition Consistency Regularization}
While the Cross-Entropy loss ($\mathcal{L}_{\text{CE}}$) optimizes the final item retrieval accuracy, it provides only implicit supervision to the internal semantic transition tensor $\mathcal{T}$. To explicitly structure the semantic space and ensure that $\mathcal{T}$ captures discriminative transition patterns, we introduce a \textit{Transition Consistency} auxiliary task. This objective enforces a pairwise ranking constraint, requiring the semantic transition score of a ground-truth successor to be strictly higher than that of unrelated items.

Formally, for each positive transition $(v_t, v_{t+1})$ in the training sequence, we contrast it against negative items $v^-$ sampled from the current mini-batch $\mathcal{B}$. The regularization loss is formulated as:
\begin{equation}
\mathcal{L}_{\text{Trans}} = - \sum_{(v_t, v_{t+1}) \in \mathcal{S}} \mathbb{E}_{v^- \sim P_{\mathcal{B}}} \left[ \log \sigma \left( T(v_t, v_{t+1}) - T(v_t, v^-) \right) \right],
\end{equation}
where $\sigma(\cdot)$ denotes the sigmoid function, and $T(\cdot, \cdot)$ represents the semantic transition score derived from $\mathcal{T}$ (as defined in Eq.~\eqref{eq:trans-score}). Here, $P_{\mathcal{B}}$ represents the in-batch negative distribution, defined as a uniform distribution over all items in the mini-batch $\mathcal{B}$ excluding the target item. This strategy ensures computational efficiency by reusing in-batch computations for negative sampling.

\subsubsection{Overall Objective}
The final objective function is a weighted sum of the next-item prediction loss and the auxiliary transition regularization:
\begin{equation}
\label{eq:total-loss}
\mathcal{L} = \mathcal{L}_{\text{CE}} + \gamma \mathcal{L}_{\text{Trans}},
\end{equation}
where $\gamma$ is a hyper-parameter controlling the weight of the transition regularization. This joint optimization ensures that while the model learns to predict user behaviors via $\mathcal{L}_{\text{CE}}$, the learnable transition tensor $\mathcal{T}$ is simultaneously supervised by $\mathcal{L}_{\text{Trans}}$ to respect item-to-item complementary priors. In scenarios with sparse or noisy interactions, this dual supervision prevents the model from overfitting to accidental co-occurrences by maintaining semantic consistency in the transition space.

\section{Experiments}
We conduct extensive experiments to evaluate the performance of \textbf{CAST}. Specifically, we focus on the following research questions:

\begin{itemize}
    \item [\textbf{RQ1:}] How does CAST perform compared against state-of-the-art SR baselines regarding accuracy and efficiency?
    \item [\textbf{RQ2:}] What is the impact of the individual components of CAST on the model's performance?
    \item [\textbf{RQ3:}] How do hyper-parameter settings and the selection of pre-trained language models affect the performance of CAST?
    \item [\textbf{RQ4:}] Can CAST effectively distinguish between complementary item pairs and random ones?
\end{itemize}

\subsection{Experimental Settings}

\subsubsection{Datasets.}

To evaluate the performance of the CAST framework, we conduct experiments on three category-based datasets derived from the Amazon Review dataset\footnote{\href{https://huggingface.co/datasets/McAuley-Lab/Amazon-Reviews-2023}{https://huggingface.co/datasets/McAuley-Lab/Amazon-Reviews-2023}}: \textit{Industrial}, \textit{Office}, and \textit{Baby}. We select these domains because they exhibit rich functional complementary relations driven by objective compatibility -- a proper setting for validating semantic complementarity. Following established protocols~\cite{kang2018self, du2023frequency}, we adopt the 5-core setting for filtering users and items with fewer than five interactions. Consistent with recent work~\cite{xu2023towards,hou2023learning,xu2024sequence}, we construct textual features by concatenating \textit{titles}, \textit{categories}, and \textit{brands}. The statistics of the datasets are summarized in Table~\ref{tab:datasets}.

\begin{table}[h]
\centering
\caption{Statistics of processed datasets.}
\label{tab:datasets}
\setlength\tabcolsep{3.2pt}
%\resizebox{\linewidth}{!}{
\begin{tabular}{lccccc}
\hline
\textbf{Dataset} & \textbf{\#Users} & \textbf{\#Items} & \textbf{\#Interactions} & \textbf{Sparsity} & \textbf{Avg.len} \\ \hline
Industrial & 50,985 & 25,848 & 361,962 & 99.972\% & 7.10 \\
Office & 223,308 & 77,551 & 1,577,570 & 99.991\% & 7.07 \\
Baby & 150,777 & 36,013 & 1,090,306 & 99.977\% & 8.23 \\ \hline
\end{tabular}
\end{table}

\subsubsection{Baseline Methods.}
To evaluate the effectiveness of our method, we compare it with \textit{ID-based SRs}: SASRec~\cite{kang2018self}, BERT4Rec~\cite{sun2019bert4rec}, SINE~\cite{tan2021sparse}, CORE~\cite{hou2022core}, CL4SRec~\cite{xie2022contrastive}, DuoRec~\cite{qiu2022contrastive}, FEARec~\cite{du2023frequency}, SASRecCPR~\cite{chang2024copy},  \textit{text-enhanced SRs}: TedRec~\cite{xu2024sequence}, and \textit{semantic ID-based SRs}: VQRec~\cite{hou2023learning}, CCFRec~\cite{liu2025bridging}.

\subsubsection{Evaluation Metrics.}
We employ the leave-one-out strategy~\cite{du2023frequency,hou2023learning}, using the last and second-to-last interactions for testing and validation, respectively, while utilizing the remaining items for training. To ensure an unbiased evaluation, we follow the protocol used in~\cite{du2023frequency,krichene2020sampled,hou2023learning} by performing the full-ranking evaluation, which ranks the ground-truth item across the entire item set. The performance is reported using standard metrics~\cite{xu2024sequence,liu2025bridging}: Recall@$K$ and NDCG@$K$, where $K \in \{5, 10, 20\}$.

\subsubsection{Implementation Details and Parameter Settings.}
\label{sec:implementation_details}
For fair comparison and reproducibility, we implement our proposed model and all baselines using the open-source recommendation library RecBole~\footnote{\url{https://github.com/RUCAIBox/RecBole}} \cite{xu2023towards}. All experiments are conducted on the Linux server equipped with one NVIDIA A100 GPU (40GB). The maximum sequence length is set to 20. For the complementary relation construction in Section \ref{sec:comp-construction}, we set the sliding window size $w=3$ and the frequency threshold $\theta_f=2$ to capture co-purchase patterns while filtering out accidental co-occurrences. The LLM confidence threshold $\theta_c=0.5$ is empirically selected to filter out low-confidence pairs while maintaining sufficient coverage of complementary relations. We set the textual similarity threshold $\theta_s=0.85$ to identify substitutable items with high semantic overlap. Optimization is done with Adam optimizer learning rate of 0.001, batch size 1024, and early stopping with a patience of 10 epochs based on the validation set's NDCG@10. The code for all the baselines is publicly available~\footnote{\url{https://github.com/Nishikata97/Baseline_RecBole}}. For ID-based baselines, we follow the optimal hyper-parameters from RecBole.

To ensure a fair comparison, we unify the backbone architecture for all Transformer-based models (including text-enhanced baselines and our proposed model) to 2 Transformer layers, 2 attention heads, and the hidden dimension of $d = 128$, and the FFN inner dimension of 256. The number of semantic subspaces $D=32$ and the codebook size $C=256$. The transition strength is set to $\lambda = 1.2$, and the regularization weight is set to $\gamma = 1.0$. The alignment dropout rate $\eta$ is set to 0.2 for Industrial and Office datasets, and 0.1 for the Baby dataset. The temperature $\tau$ is set to 0.07. Furthermore, we use \texttt{Qwen3-4B}~\cite{yang2025qwen3} as the PLM encoder for textual feature extraction, where the text embeddings are reduced to $d_{\text{text}} = 128$ dimensions using PCA, and the Faiss library \cite{johnson2019billion} for semantic ID generation. We employ \texttt{gemma-3-27b-it}~\cite{team2025gemma} for complementary relation inference.

\begin{table*}[t]
\centering
\caption{Overall performance comparison. The best  performances are highlighted in \textbf{bold}, and the second-best performances are \underline{underlined}.}
\label{tab:overall-performance}
\setlength\tabcolsep{3.5pt}
%\resizebox{0.95\textwidth}{!}{
\begin{tabular}{ccccccccccccc}
\toprule
\textbf{Dataset} & \multicolumn{4}{c}{\textbf{Industrial}} & \multicolumn{4}{c}{\textbf{Office}} & \multicolumn{4}{c}{\textbf{Baby}} \\
\cmidrule(lr){2-5} \cmidrule(lr){6-9} \cmidrule(lr){10-13}
\multirow{2}{*}{\textbf{Metrics}} 
& \multicolumn{2}{c}{\textbf{Recall}} & \multicolumn{2}{c}{\textbf{NDCG}} 
& \multicolumn{2}{c}{\textbf{Recall}} & \multicolumn{2}{c}{\textbf{NDCG}} 
& \multicolumn{2}{c}{\textbf{Recall}} & \multicolumn{2}{c}{\textbf{NDCG}} \\
& @5 & @10 & @5 & @10 & @5 & @10 & @5 & @10 & @5 & @10 & @5 & @10 \\
\midrule
SASRec & 2.45 & 3.90 & 1.38 & 1.84 & 2.56 & 3.83 & 1.49 & 1.90 & 2.27 & 3.66 & 1.39 & 1.84 \\
BERT4Rec  & 1.64 & 2.73 & 1.05 & 1.40 & 1.67 & 2.59 & 1.10 & 1.39 & 1.69 & 2.80 & 1.10 & 1.45 \\
SINE & 2.27 & 3.60 & 1.47 & 1.90 & 2.17 & 3.27 & 1.42 & 1.77 & 1.90 & 3.10 & 1.21 & 1.59 \\
CORE & 1.54 & 3.27 & 0.77 & 1.33 & 1.90 & 3.20 & 1.01 & 1.43 & 1.61 & 2.84 & 0.92 & 1.32 \\
CL4SRec & 2.65 & 4.20 & 1.66 & 2.16 & 2.55 & 3.75 & 1.65 & 2.04 & 2.37 & 3.79 & 1.47 & 1.93 \\
DuoRec & 2.28 & 3.54 & 1.34 & 1.75 & 2.36 & 3.48 & 1.40 & 1.76 & 2.14 & 3.44 & 1.27 & 1.68 \\
FEARec & 2.51 & 3.87 & 1.43 & 1.86 & 2.59 & 3.80 & 1.50 & 1.89 & 2.25 & 3.64 & 1.36 & 1.81 \\
SASRecCPR & 2.18 & 3.45 & 1.43 & 1.83 & 2.15 & 3.11 & 1.49 & 1.80 & 2.11 & 3.41 & 1.39 & 1.80 \\
\midrule
\multicolumn{13}{c}{\textbf{Text-enhanced SRs}} \\
\midrule
TedRec & 2.58 & 4.01 & 1.67 & 2.13 & 2.51 & 3.78 & 1.62 & 2.03 & 2.20 & 3.58 & 1.36 & 1.80 \\
VQRec & 2.81 & 4.40 & 1.50 & 2.02 & 2.30 & 3.39 & 1.51 & 1.86 & 2.12 & 3.40 & 1.30 & 1.71 \\
CCFRec & \underline{3.18} & \underline{5.10} & \underline{1.93} & \underline{2.55} & \underline{2.68} & \underline{4.02} & \underline{1.75} & \underline{2.18} & \underline{2.68} & \underline{4.24} & \underline{1.71} & \underline{2.21} \\
\midrule
\textbf{CAST} & \textbf{3.45} & \textbf{5.45} & \textbf{2.16} & \textbf{2.81} & \textbf{3.14} & \textbf{4.73} & \textbf{2.02} & \textbf{2.53} & \textbf{2.77} & \textbf{4.42} & \textbf{1.79} & \textbf{2.31} \\
\small \textit{Improv.} & \small 8.49\% & \small 6.86\% & \small 11.92\% & \small 10.20\% & \small 17.16\% & \small 17.66\% & \small 15.43\% & \small 16.06\% & \small 3.36\% & \small 4.25\% & \small 4.68\% & \small 4.50\% \\
\bottomrule
\end{tabular}
\end{table*}

\subsection{Overall Performance (RQ1)}

Table~\ref{tab:overall-performance} provides a comparative analysis of CAST against all the baselines evaluated across three real-world datasets with relative improvements (Improv. (\%)) of CAST's performance over the strongest baseline. We make the following observations.

First, within traditional ID-based methods (e.g., SASRec, SINE, CORE, FEARec, and SASRecCPR), FEARec and SASRec consistently achieve slightly better performance than the others. SASRec utilizes a self-attention encoder to model both short-term and long-term item transitions. FEARec builds on SASRec by integrating hybrid time–frequency attention to capture both high- and low-frequency patterns in user sequences. In contrast, SINE emphasizes multi-interest retrieval and representation consistency, while CORE and SASRecCPR, respectively, focus on output-layer adjustments and duplicate-item copying mechanisms, which appear to have less advantage under our full-ranking evaluation setting.

Second, text-enhanced SR methods (TedRec, VQRec, and CCFRec) generally outperform the above ID-centric methods by leveraging item textual information as richer semantic carriers. TedRec and VQRec both surpass SASRec and FEARec on most metrics, and CCFRec stands out as the strongest baseline overall, demonstrating the clear advantage of representing items in a semantic space rather than relying solely on discrete IDs. However, these methods still keep semantics in a continuous, model-specific latent space, making it difficult to interpret why two items should be complementary and limiting the ability to explicitly model fine-grained semantic transitions that underlie cross-item complementarity.

Finally, CAST achieves the best performance across all datasets and evaluation metrics. Compared with the strongest baseline CCFRec, our model yields consistent relative improvements ranging from 3.36\% to 17.66\% on Recall and NDCG.

\subsection{Ablation Study (RQ2)}

To analyze the individual contribution of CAST's components, we compare our full model with three variants that remove key modules (Table~\ref{tab:ablation}).

\textit{(1) w/o Semantic Codes (Text-Only)}: We remove the semantic ID embedding modeling and directly feed PLM-derived text embeddings into the sequential model; this aims to verify whether the discrete semantic IDs provide discriminative information beyond raw textual representations.

\textit{(2) w/o Alignment}: We replace the subspace-preserving semantic-text alignment (Section~\ref{sec:alignment}) with a simple mean pooling operation over semantic code embeddings, followed by element-wise addition to text embeddings; this examines the necessity of the proposed non-linear fusion and subspace preservation.

\textit{(3) w/o Trans. Guide}: We disable the transition modeling by setting both the relation weight $\lambda=0$ in Eq.~\eqref{eq:trans-attn} and the regularization weight $\gamma=0$ in Eq.~\eqref{eq:total-loss}; this simplifies the encoder to a standard self-attention mechanism, validating the importance of incorporating semantic transfer signals.

%\textit{(4) w/o LLM-inference}: We construct relations based on co-purchase statistics, removing LLM filtering and transitive expansion, thereby validating the effectiveness of the LLM-based refinement strategy in Algorithm~\ref{algorithm:construction} by showing that co-occurrence signals differ significantly from complementary relations.

As shown in Table~\ref{tab:ablation}, removing any component leads to performance degradation across all datasets. Specifically, removing semantic codes (\textit{w/o Sem. Codes}) leads to the most severe performance degradation, indicating that OPQ-based semantic IDs are essential for capturing high-level item semantics beyond textual similarity. \textit{w/o Trans. Guide} also shows a consistent decrease, suggesting that the transition $\lambda T_{i,j}$ incorporates global semantic-level transition priors, which are crucial for guiding predictions when the standard self-attention is dominated by semantic similarity under data sparsity. Finally, \textit{w/o Alignment} underperforms the full model, highlighting that our non-linear projection effectively bridges the semantic gap between codes and text, whereas simple pooling leads to information loss.

\begin{table}[t]
\centering
\caption{Ablation analysis of key components in CAST.}
\label{tab:ablation}
\setlength\tabcolsep{2.6pt}
\begin{tabular}{lcccccc}
\toprule
\textbf{Dataset} 
& \multicolumn{2}{c}{\textbf{Industrial}} 
& \multicolumn{2}{c}{\textbf{Office}} 
& \multicolumn{2}{c}{\textbf{Baby}} \\
\cmidrule(lr){2-3} \cmidrule(lr){4-5} \cmidrule(lr){6-7}
\textbf{Metrics} 
& R@10 & N@10 
& R@10 & N@10 
& R@10 & N@10 \\
\midrule
(0) \textbf{CAST}             & \textbf{5.45} & \textbf{2.81} & \textbf{4.73} & \textbf{2.53} & \textbf{4.42} & \textbf{2.31} \\
(1) \textit{w/o} Sem. Codes   & 3.91 & 1.99 & 2.72 & 1.51 & 2.87 & 1.48 \\
(2) \textit{w/o} Alignment    & 4.87 & 2.49 & 4.32 & 2.31 & 4.19 & 2.18 \\
(3) \textit{w/o} Trans. Guide & 4.99 & 2.58 & 4.45 & 2.38 & 4.29 & 2.24 \\
\bottomrule
\end{tabular}
\end{table}
%(4) \textit{w/o} LLM-infer    & 0.00 & 0.00 & 0.00 & 0.00 & 0.00 & 0.00 \\

\subsection{Impact of Hyper-parameters (RQ3)}

%In this section, we investigate the impact of key hyper-parameters and component choices on CAST's performance, including the selection of the pre-trained language model (PLM).

\begin{figure*}[h]
  \centering
  \includegraphics[width=0.75\textwidth]{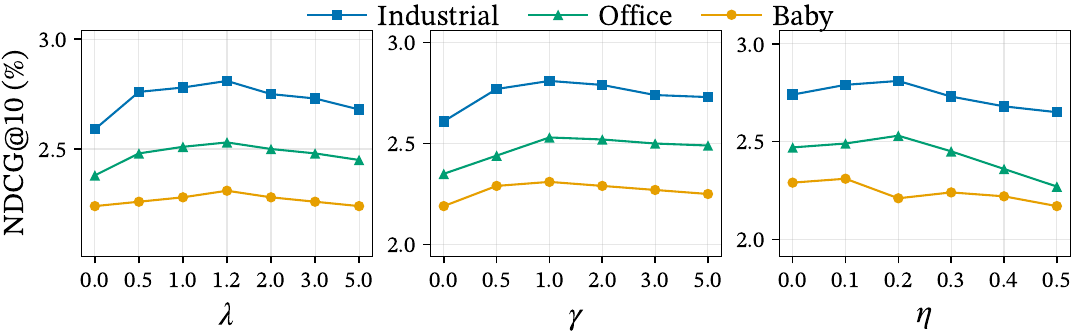}
  \caption{Hyper-parameter analysis of CAST on three datasets in terms of NDCG@10.}
  \label{fig:hyperparam}
\end{figure*}

We investigate the impact of semantic transition strength $\lambda$, transition consistency loss weight $\gamma$, and alignment dropout probability $\eta$ on model performance (Figure~\ref{fig:hyperparam}). Additionally, the influence of different PLM encoders is explored in Section~\ref{sec:plm_comparison}.

\subsubsection{Impact of Semantic Transition Strength ($\lambda$)}
The parameter $\lambda$ balances the contribution of the global semantic transition scores against the self-attention scores (Eq.~\eqref{eq:trans-attn}). The initial improvement (peaking at $\lambda=1.2$) validates that injecting transition scores effectively alleviates the sparsity issue of pure interaction data. However, the degradation observed when $\lambda > 2.0$ indicates that a large $\lambda$ forces the model to over-rely on global transition patterns, thereby suppressing personalized user preferences.

\subsubsection{Impact of Transition Regularization ($\gamma$)}
The parameter $\gamma$ controls the weight of the transition consistency loss $\mathcal{L}_{\text{Trans}}$ in the overall objective (Eq.~\eqref{eq:total-loss}). The lower performance at $\gamma=0$ suggests that despite the prior initialization, without explicit supervision, the learnable transition tensor $\mathcal{T}$ may fail to retain meaningful complementary dependencies during the optimization process. The performance remains robust within $\gamma \in [1.0, 2.0]$, validating that enforcing transition consistency via $\mathcal{L}_{\text{Trans}}$ is essential for learning reliable semantic relations.

\subsubsection{Impact of Alignment Dropout ($\eta$)}
The parameter $\eta$ controls the dropout probability in the subspace-preserving alignment module (Eq.~\eqref{eq:nonlinear}). The results show that moderate regularization (peaking at $\eta = 0.2$ for most datasets) yields the optimal performance, balancing generalization and feature preservation. However, a large dropout rate ($\eta \ge 0.4$) degrades model performance, indicating that the alignment between fine-grained semantic codes and textual embeddings is sensitive to information loss.

\subsubsection{Comparison of Pre-trained Language Models}
\label{sec:plm_comparison}
We investigate the performance of the model under different pre-trained language models used to generate semantic codes and semantic-text alignment in Section~\ref{sec:alignment}. Specifically, as shown in Table~\ref{tab:plm}, we compare three representative models: T5-base~\cite{raffel2020exploring}, T5-xxl~\cite{tay2022scale}, and Qwen3-4B~\cite{yang2025qwen3}. Empirically, Qwen3-4B delivers the best performance across all evaluation metrics, highlighting the advantage of high-quality, large-scale pre-training corpora in modeling fine-grained semantics. Although T5-xxl has larger parameters than T5-base, we observe that both models yield comparable performance.

\begin{table}[t]
\centering
\caption{Comparison of different pre-trained language models for semantic code generation and alignment.}
\label{tab:plm}
\setlength\tabcolsep{3.5pt}
\begin{tabular}{lcccccc}
\toprule
\textbf{Dataset} & \multicolumn{2}{c}{\textbf{Industrial}} & \multicolumn{2}{c}{\textbf{Office}} & \multicolumn{2}{c}{\textbf{Baby}} \\
\cmidrule(lr){2-3} \cmidrule(lr){4-5} \cmidrule(lr){6-7}
\textbf{Model} & \textbf{R@10} & \textbf{N@10} & \textbf{R@10} & \textbf{N@10} & \textbf{R@10} & \textbf{N@10} \\
\midrule
T5-base & 5.36 & 2.75 & 4.46 & 2.41 & 4.32 & 2.24 \\
T5-xxl & 5.21 & 2.67 & 4.52 & 2.44 & 4.32 & 2.23 \\
Qwen3-4B & \textbf{5.45} & \textbf{2.81} & \textbf{4.73} & \textbf{2.53} & \textbf{4.42} & \textbf{2.31} \\
\bottomrule
\end{tabular}
\end{table}

\subsection{Efficiency Comparison (RQ1)}
\label{sec:efficiency}

Table~\ref{tab:efficiency} reports the average training time per epoch (\textit{Time}) and maximum GPU memory usage (\textit{GPU memory}) on the Office dataset using a single NVIDIA A100-40G. In contrast to CCFRec, which is restricted to a batch size of 128 due to memory constraints caused by its heavy architecture, CAST trains efficiently with standard batch sizes (e.g., 1024) on the same hardware. Consequently, it yields a 65.05$\times$ acceleration in training time and a 9.26$\times$ reduction in memory usage compared to CCFRec, while maintaining superior Recall.

\textbf{Complexity Analysis.} The observed efficiency gains are underpinned by our lightweight design. Existing semantic-based methods typically employ heavy cross-attention mechanisms (e.g., Q-Former \cite{li2023blip}) to align semantic codes with textual representations, resulting in a computational complexity of $\mathcal{O}(N^2)$ where $N$ is the number of tokens. In contrast, CAST reduces the alignment complexity to $\mathcal{O}(N)$ via subspace-preserving MLP (Section \ref{sec:alignment}) and maintains standard attention complexity $\mathcal{O}(L^2 d)$ for transition modeling (Section \ref{sec:comp-modeling}).

\begin{table}[t]
\centering
\caption{Efficiency analysis on Office. Metrics include time per epoch (s) and Maximum GPU memory (MB). CCFRec$^\dag$ is capped at batch size 128 by hardware limits.}
\label{tab:efficiency}
\begin{tabular}{llll}
\hline
\textbf{Model}                      & \textbf{Time} & \textbf{GPU Memory} & \textbf{R@10} \\ \hline
SASRec                     & 34                               & 2,480                        & 3.83               \\
FEARec                     & 59                               & 2,579                        & 3.80               \\
CL4SRec                    & 373                              & 3,114                        & 3.75               \\
TedRec                     & 60                               & 3,384                        & 3.79               \\
CCFRec$^\dagger$                    & 4944                              & 35,803                       & 4.51 \\ 
\midrule % \hline
\textbf{CAST}                & \textbf{76}                      & \textbf{3,866}               & \textbf{4.73}      \\ \hline
\end{tabular}
\end{table}

\subsection{Analysis of Complementary Transition Learning (RQ4)}

\begin{figure}[t]
  \centering
  \includegraphics[width=0.8\linewidth]{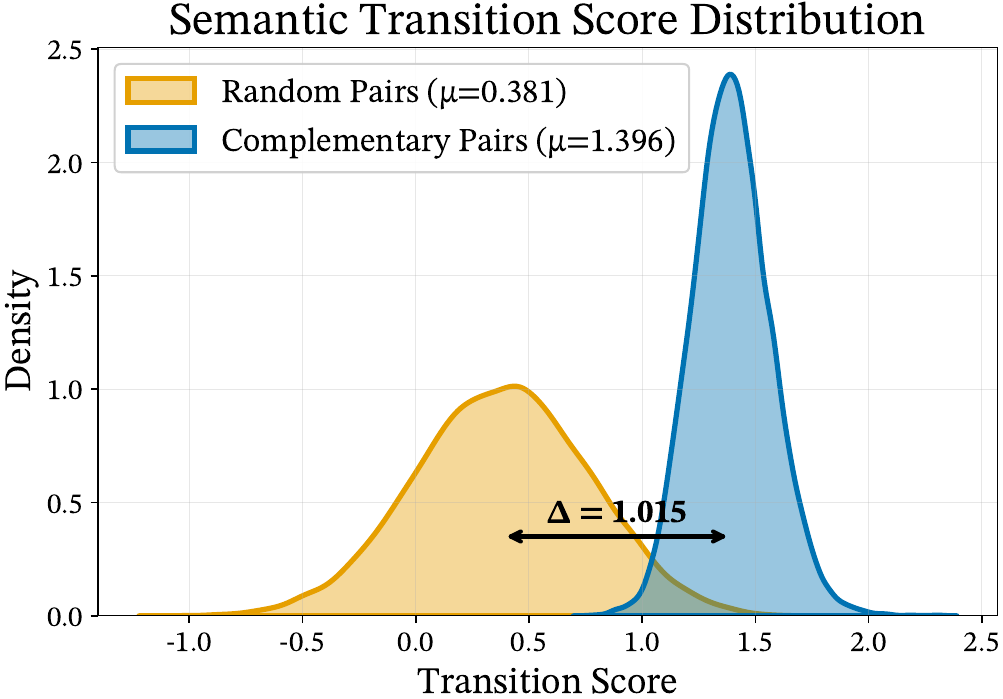}
  \caption{Distribution of semantic transition scores.}
  \label{fig:transition}
\end{figure}

To analyze whether semantic transition scores differentiate complementary item relations from unrelated pairs, Figure~\ref{fig:transition} compares the score distributions for complementary item pairs and randomly sampled pairs. Complementary pairs exhibit substantially higher scores than random pairs ($\mu_{\text{comp}} = 1.396$ vs.\ $\mu_{\text{rand}} = 0.381$, $\Delta = 1.015$), with only limited overlap between the two distributions. This observation indicates that transition scores significantly differ between complementary and unrelated item pairs, reflecting the structured semantic dependencies captured by the transition module in Section \ref{sec:comp-modeling}.

\section{Related Work}

\subsection{ID-based Sequential Recommendation}
Early methods~\cite{rendle2010factorizing} integrated Markov chains with Matrix Factorization to model item-to-item transitions. Development of deep learning introduced neural network approaches, including CNNs~\cite{tang2018personalized}, RNNs~\cite{hidasi2016session,li2017neural}, and GNNs~\cite{chang2021sequential}. Recently, Transformer-based architectures have been widely adopted as the backbone for sequential recommendation. SASRec~\cite{kang2018self} and BERT4Rec~\cite{sun2019bert4rec} leverage self-attention mechanisms to model long-range dependencies, while more recent variants, like FEARec~\cite{du2023frequency}, perform sequence modeling in the frequency domain to mitigate noise in time-domain interactions. Despite their success, these methods fundamentally rely on atomic \textit{item IDs}, treating items as independent tokens. Consequently, item ID-based methods suffer from the cold-start problem and data sparsity, as they fail to capture the latent semantic reasons for transitions and rely on historical co-occurrence statistics.

\subsection{Text- and Semantic-based Sequential Recommendation}
To overcome the limitations of item ID-based methods, researchers have explored learning semantic representations directly from item text. Early works like UniSRec \cite{hou2022towards} utilize PLMs to encode item text into universal representations. However, such continuous representations may over-emphasize text features, limiting the capture of collaborative signals. To address this, VQ-Rec \cite{hou2023learning} and CCFRec \cite{liu2025bridging} introduce discrete semantic IDs derived from text, decoupling text encoding from representation learning. However, a critical limitation of these semantic ID-based methods lies in their coarse-grained sequence modeling. Despite utilizing discrete semantic codes, they essentially aggregate these fine-grained semantics (e.g., via mean pooling or attention fusion) into a unified item representation before sequence modeling. This \textit{aggregation process} obscures the sequential dependencies between specific semantic attributes, forcing the model to capture transitions solely at the item-level. CAST models transitions directly in the semantic code space, capturing the dependencies of item semantics during sequence modeling.

\subsection{Complementary Modeling in Recommendation}
Identifying complementary items (e.g., camera and lens) is crucial for capturing diverse user needs beyond simple similarity~\cite{mcauley2015inferring}. While early works~\cite{wang2020make,zhang2021learning} utilized relation graphs to model these dependencies, they relied on co-purchase statistics, which often conflate true complementary relations with spurious correlations (e.g., popularity bias)~\cite{sugahara2024really}. Recent advances in LLMs allow us to leverage extensive world knowledge to infer item relationships. LLMRec~\cite{wei2024llmrec} utilizes LLMs to denoise interaction graphs and infer missing edges, while other recent works~\cite{li2024explainable} employ LLMs to generate explainable complementary recommendations directly. However, these methods typically treat LLM knowledge as static external signals, disconnected from the dynamic sequence modeling process. CAST addresses this by internalizing LLM-verified complementary priors into learnable semantic-level transitions. This allows the model to adaptively refine functional compatibility patterns through user interactions, effectively bridging the gap between static relations and dynamic user preferences.

\section{Conclusion}
This work presents CAST, a framework for complementary-aware sequential recommendation that models fine-grained semantic-level transitions. Our findings demonstrate that explicitly modeling dynamic dependencies in a discrete semantic space effectively uncovers latent functional compatibility, which is often obscured in traditional ID-based approaches. By modeling latent semantic dependencies, CAST can distinguish true complementary patterns from spurious correlations. Despite these advancements, two limitations remain. First, while the functional complementarity assumption fits e-commerce, it may be less effective in content-consumption domains driven by topical similarity. Second, our semantic codes remain latent, lacking direct mapping to explicit attributes (e.g., specifications) for explainability. Future work will explore (1) \textit{Adaptive Multi-Intent Modeling} to generalize across domains by balancing functional and topical information, and (2) \textit{Interpretable Attribute Alignment} to map latent codes to concrete item features.

%%
%% The next two lines define the bibliography style to be used, and
%% the bibliography file.
\bibliographystyle{ACM-Reference-Format}
\bibliography{main-base}

%%
%% If your work has an appendix, this is the place to put it.
%\appendix
%\section{Section 1 in Appendix}

\end{document}